\documentclass[twocolumn,apl,epsf,showpacs,superscriptaddress]{revtex4}
\usepackage{amsmath}
\usepackage{amsfonts}
\usepackage{epsfig}
\usepackage{epsf}
\usepackage{array}

\begin{document}

\title{Spin injection and spin transport in paramagnetic insulators}
\altaffiliation{Copyright  notice: This  manuscript  has  been  authored  by  UT-Battelle, LLC under Contract No. DE-AC05-00OR22725 with the U.S.  Department  of  Energy.   
The  United  States  Government  retains  and  the  publisher,  by  accepting  the  article  for  publication, 
acknowledges  that  the  United  States  Government  retains  a  non-exclusive, paid-up, irrevocable, world-wide license to publish or reproduce the published form of this manuscript, 
or allow others to do so, for United States Government purposes.  
The Department of Energy will provide public access to these results of federally sponsored  research  in  accordance  with  the  DOE  Public  Access  Plan 
(http://energy.gov/downloads/doe-public-access-plan)}
\author{Satoshi Okamoto}
\altaffiliation{okapon@ornl.gov}
\affiliation{Materials Science and Technology Division, Oak Ridge National Laboratory, Oak Ridge, Tennessee 37831, USA}

\begin{abstract}
We investigate the spin injection and the spin transport in paramagnetic insulators described by simple Heisenberg interactions using auxiliary particle methods. 
Some of these methods allow access to both paramagnetic states above magnetic transition temperatures and magnetic states at low temperatures. 
It is predicted that the spin injection at an interface with a normal metal is rather insensitive to temperatures above the magnetic transition temperature. 
On the other hand below the transition temperature, 
it decreases monotonically and disappears at zero temperature. 
We also analyze the bulk spin conductance. 
It is shown that the conductance becomes zero at zero temperature as predicted by linear spin wave theory 
but increases with temperature and is maximized around the magnetic transition temperature. 
These findings suggest that the compromise between the two effects determines 
the optimal temperature for spintronics applications utilizing magnetic insulators. 
\end{abstract}


\maketitle

\date{\today }


\section{Introduction}

Manipulating spin degrees of freedom of electrons in addition to charge degrees is at the forefront of current materials science \cite{Maekawa2006}. 
By utilizing the spin Hall effect (SHE) \cite{Dyakonov1971,Hirsch1999,Zhang2000,Murakami2003,Sinova2004}
and the inverse spin Hall effect  (ISHE)\cite{Saitoh2006}, 
a variety of novel phenomena are envisioned \cite{Bauer2012}. 
For example, the spin current in magnetic materials can be generated by the temperature gradient, the spin Seebeck effect (SSE) \cite{Uchida2008}. 
Unlike the conventional Seebeck effect, the SSE also exists in magnetic insulators (MIs) \cite{Uchida2010} 
in which the spin current is mediated by spin waves or magnons.  
Not only in a ferromagnetic (FM) insulator or a ferrimagnetic insulator, 
the spin transport is possible through an antiferromagnetic (AFM) insulator \cite{Wang2014}. 
Further, magnetic excitations can be used for the transmission of the electrical signal through MIs \cite{Kajiwara2010}. 

Conventionally, these phenomena have been analyzed based on excitations or fluctuations from magnetic ordering \cite{Maekawa13}.  
However, recent studies demonstrate such phenomena even above magnetic transition temperatures:  
spin current generation \cite{Shiomi14,Qiu15} and the SSE  \cite{Wu2015}. 
Thus, a new or extended concept which does not require magnetic ordering needs to be developed. 
Once this is established, it would extend the current scope of spintronics, allowing for example 
the transmission of electrical signals through {\em paramagnetic insulators} (PM insulators or PMIs) 
mediated by some form of magnetic fluctuations, which is appealing for both fundamental science and applications 
because it allows the transport of magnetic quanta for efficient information technologies. 

In this paper, we provide a simple description of the performance of a PMI/nomal metal (NM) interface 
by evaluating the spin current through the interface and the spin current conductance inside the PMI  using auxiliary particle (AP) methods, 
some of which allow access to both PM and magnetic states. 
We find that the spin current injection at a PMI/NM interface is sensitive to temperature below a magnetic transition temperature $T_c$ 
[$T_C$ for ferromagnet (FM) and $T_N$ for AFM]
but weakly dependent on temperatures above $T_c$. 
The spin conductance due to magnetic fluctuations is found to be maximized around $T_c$. 
These findings suggest that a MI could function as a spintronics material even above its magnetic ordering temperature. 

\section{Model and method}
We consider MIs described by the Heisenberg model with the nearest neighbor (NN) exchange coupling $J>0$ and spin $S$, 
\begin{equation}
H=\mp J\sum_{\langle \vec r \vec r' \rangle} \vec S_{\vec r} \cdot \vec S_{\vec r'}, 
\label{eq:HJ}
\end{equation}
with the negative sign for a FM and the positive sign for an AFM. 
In order to describe MIs below and above their magnetic transition temperature, 
we employ AP methods, Schwinger boson (SB) as well as Schwinger fermion (SF) mean-field (MF) methods \cite{Arovas1988}. 
Our main focus is on the finite temperature properties of a simple model, not exotic properties at low temperatures such as spin liquids. 
Thus, we consider the simplest {\it Ans{\"a}tze} for MF wave functions which do not break the underlying lattice symmetry.
These AP methods are suitable for magnetically disordered states because MF order parameters are defined on ``bonds,'' not on ``sites.'' 
Further, SB methods can also deal with magnetically ordered states because condensed SBs correspond to the ordered moments. 

Here, we briefly review AP MF methods. 
In these methods, spin operators are expressed in terms of APs as 
$S_{\vec r}^z= a^\dag_{\vec r \uparrow} a_{\vec r \uparrow} -  a^\dag_{\vec r \downarrow} a_{\vec r \downarrow}$, 
$S_{\vec r}^+= a^\dag_{\vec r \uparrow} a_{\vec r \downarrow}$ and 
$S_{\vec r}^-= a^\dag_{\vec r \downarrow} a_{\vec r \uparrow}$, 
with $a^\dag_{\vec r \sigma}(a_{\vec r \sigma})$ being the creation (annihilation) operator of an AP with spin $\sigma$ at position $\vec r$. 
Using these expressions and operator identities, the spin exchange term in Eq.~(\ref{eq:HJ}) is written as 
$\vec S_{\vec r} \cdot \vec S_{\vec r'} = \frac{1}{2} \chi_{\vec r \vec r'}^\dag \chi_{\vec r \vec r'} = - \frac{1}{2} \Delta_{\vec r \vec r'}^\dag \Delta_{\vec r \vec r'}$ for Schwinger bosons and 
$- \frac{1}{2} \chi_{\vec r \vec r'}^\dag \chi_{\vec r \vec r'}$ for Schwinger fermions \cite{Arovas1988}. 
Here, constant terms are neglected, and $\chi_{\vec r \vec r'}=\sum_\sigma a_{\vec r \sigma}^\dag a_{\vec r' \sigma}$ and  
$\Delta_{\vec r \vec r'} = a_{\vec r \uparrow} a_{\vec r' \downarrow} -a_{\vec r \downarrow} a_{\vec r' \uparrow}$ 
represent short-range magnetic correlations. 
Then, the MF decoupling is introduced as 
$-A^\dag_{\vec r \vec r'} A_{\vec r \vec r'} \Rightarrow -\langle A_{\vec r \vec r'}^\dag \rangle A_{\vec r \vec r'} -  A^\dag_{\vec r \vec r'} \langle A_{\vec r \vec r'} \rangle$. 
Thus, the negative sign in front of the operator product is essential, distinguishing the different interaction, FM or AFM, and the statistics of the AP, boson or fermion. 
Consequently, the SB MF method for FM and the SF MF method for AFM can be formulated in parallel, and the MF Hamiltonian is given by 
\begin{equation}
H_{FSB/ASF} = -\frac{J\chi}{2}\!\! \sum_{\vec r, \hat \rho, \sigma} \!\! \bigl( a_{\vec r \sigma}^\dag a_{\vec r+\hat \rho \sigma} + {\rm H.c.} \bigr) + 
\lambda \sum_{\vec r, \sigma} a_{\vec r \sigma}^\dag a_{\vec r \sigma}. 
\end{equation}
Here, $\hat \rho$ is a unit vector connecting a nearest neighbor site in the positive direction, and constant terms are neglected. 
$\lambda$ is the Lagrange multiplier enforcing the constraint for APs,  
$\sum_{\sigma} \langle a_{\vec r \sigma}^\dag a_{\vec r \sigma}  \rangle = 2S$ (for SFs, $S=1/2$), 
and $\chi=\langle \chi_{\vec r \vec r+\hat \rho} \rangle$ is the bond order parameter. 
This MF Hamiltonian is diagonalized to yield the dispersion relation for APs as 
$\omega_{\vec q \sigma} = \lambda - J \chi \gamma_{\vec q}$, where 
$\gamma_{\vec q} = \sum_{\rho} \cos q_\rho $. 
Similarly, the SB MF Hamiltonian for an AFM is given by 
\begin{eqnarray}
H_{ASB} \!\!&=& \!\! -\frac{J \Delta}{2} \sum_{\vec r, \hat \rho} 
\bigl( a_{\vec r \uparrow} a_{\vec r + \hat \rho \downarrow} -a_{\vec r \downarrow} a_{\vec r + \hat \rho \uparrow} + {\rm H.c.} \bigr) \nonumber \\ 
&&+ 
\lambda \sum_{\vec r, \sigma} a_{\vec r \sigma}^\dag a_{\vec r \sigma}. 
\end{eqnarray}
Here, the MF order parameter is 
$\Delta = \langle \Delta_{\vec r \vec r + \hat \rho} \rangle$. 
The excitation spectrum is given by $\omega_{\vec q} = \sqrt{\lambda^2 - | J \Delta \gamma'_{\vec q}|^2}$ 
with $\gamma'_{\vec q} = i \sum_{\rho} \sin q_\rho$. 

Since we are dealing with non-interacting particles under the MF approximation, 
it is straightforward to compute the Green's functions of APs. 
The Green's function for FM SBs or AFM SFs is given by $D_{\vec q \sigma}(\tau) = - \langle T_\tau a_{\vec q \sigma}(\tau) a^\dag_{\vec q \sigma}(0) \rangle$. 
In the Matsubara frequency, $\nu_n=2n \pi T$ for SB and $(2n+1) \pi T$ for SF, this is expressed by 
\begin{equation}
D_{\vec q \sigma} (i\nu_n) = \frac{1}{i \nu_n - \omega_{\vec q}}. 
\label{eq:D1}
\end{equation}
The Green's function for AF SBs is given using the Nambu representation by 
$\hat D_{\vec q}(\tau) = - \langle T_\tau \Psi_{\vec q \sigma}(\tau) \Psi^\dag_{\vec q \sigma}(0) \rangle$
with $\Psi_{\vec q \sigma}= (a_{\vec q \uparrow}, a_{-\vec q \downarrow}^\dag)$. 
In the Matsubara frequency, this is expressed by the matrix 
\begin{eqnarray}
&&\hspace{-3em}\hat D_{\vec q}(i\nu_n) = 
\left[
\begin{array}{cc}
D_{\vec q \uparrow}(i\nu_n) & F_{\vec q}(i\nu_n) \\
F_{\vec q}^*(i \nu_n) & D_{-\vec q \downarrow}(-i\nu_n)
\end{array}
\right] \nonumber \\
&&\hspace{-1em}=\frac{1}{2 \omega_{\vec q}} \sum_{s = \pm 1}
\biggl(\frac{s}{i\nu_n-s \omega_{\vec q}}\biggr)
\left[
\begin{array}{cc}
i\nu_n + \lambda & -i J\Delta \gamma'_{\vec q} \\
i J\Delta \gamma'_{\vec q} & -i\nu_n + \lambda
\end{array}
\right] \hspace{-0.3em}. 
\label{eq:D2}
\end{eqnarray}

 In principle, MF order parameters, $\chi$ for FM SB and AFM SF and $\Delta$ for AFM SB, must be fixed by solving self consistent equations. 
However, it is known that there appear fictitious phase transitions above which order parameters disappear. 
In three-dimensional bosonic systems, this happens simultaneously with the disappearance of the boson condensation \cite{Arovas1988}. 
To avoid such fictitious phase transitions, we treat $\chi$ and $\Delta$ as constants. 
This approach may correspond to an effective mass approximation used in literature, 
but here the underling lattice with the discrete translational symmetry is maintained. 
It should also be noted that these parameters would introduce additional temperature dependence at high temperatures because 
$\chi^2$  or $\Delta^2$ roughly corresponds to $|\langle \vec S_{\vec r} \cdot \vec S_{\vec r'}\rangle| \propto T^{-1}$. 
Since we are interested in the behavior of MIs both below and above their transition temperatures, 
we focus on three-dimensional systems throughout this paper.

\begin{figure}[tbp]
\includegraphics[width=0.8\columnwidth, clip]{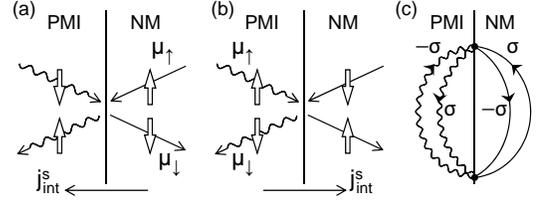}
\caption{Schematic diagram of a PMI/NM interface. 
(a) Due to the interfacial exchange coupling $H_{int}$, the electron reflection accompanied with the spin flip causes the spin flip for APs, 
resulting in interfacial spin current $j_{int}^s$. 
(b) Opposite process in which 
the reflection of APs accompanied with the spin flip causes the spin flip for electrons. 
$\mu_\uparrow - \mu_\downarrow = \delta\mu$ is the splitting in the chemical potential between up and down electrons (APs) in (a) [(b)]. 
Process (b) is also called spin extraction. 
(c) Feynman diagram representing the spin current injection. 
Solid lines are propagators for electrons, while wavy lines are for APs.
Dots represent $H_{int}$. }
\label{fig:cartoon}
\end{figure}

\section{Theoretical results and discussion}
{\em Spin-current injection at PMI/NM interfaces}.
For the interfacial magnetic interaction, we consider the following exchange coupling: 
\begin{equation}
H_{int}=J_{eff} \hspace{-0.7em} \sum_{\vec q, \vec q', \vec k, \vec k'} \hspace{-0.7em}
\bigl( a_{\vec q \uparrow}^\dag a_{\vec q' \downarrow} c_{\vec k \downarrow}^\dag c_{\vec k' \uparrow} + {\rm H. c.} \bigr) . 
\label{eq:Hint}
\end{equation}
Here, $J_{eff}$ is the coupling constant and $c_{\vec k \sigma}$ is 
the annihilation operator of an electron inside a NM with spin $\sigma$. 
The summation over the momenta $\vec q, \vec q', \vec k, \vec k'$ is taken independently 
assuming a rough interface, where the transverse components of momenta are not conserved. 
There are two cases. 
When the splitting in the chemical potential $\delta\mu =\mu_\uparrow-\mu_\downarrow$ exists for electrons in the NM region, 
this represents the electron reflection at the interface by exchanging spin with the PMI, transferring one spin quantum to the PMI region, 
Fig.~\ref{fig:cartoon} (a). 
$\delta\mu$ corresponds to the spin voltage due to the spin accumulation induced externally 
by means of, for example, the spin Hall effect \cite{Sinova2015}. 
When the splitting in the chemical potential exists for APs in the PMI instead, 
this represents the reflection of APs at the interface by exchanging spin with the NM, transferring one spin quantum to the NM region, 
Fig.~\ref{fig:cartoon} (b). 
The former corresponds to the spin injection and the latter the spin extraction as discussed in Ref. \cite{Zutic2002}. 

For concreteness, we consider the spin injection by the spin splitting for electrons in the NM region [Fig.~\ref{fig:cartoon} (a)]. 
By the equation of motion for the spin moment inside the PM region, 
the interface spin current operator is given by 
\begin{eqnarray}
\hat j_{int}^s  \!\!&=&\!\! \partial_t \sum_{\vec r, \sigma} \sigma a^\dag_{\vec r \sigma} a_{\vec r \sigma}
=-i[\sum_{\vec r, \sigma} \sigma a^\dag_{\vec r \sigma} a_{\vec r \sigma}, H_{int}] \nonumber \\
&=&\!\! -i J_{eff} \hspace{-0.7em} \sum_{\vec q, \vec q', \vec k, \vec k'} \hspace{-0.7em}
\bigl( a_{\vec q \uparrow}^\dag a_{\vec q' \downarrow} c_{\vec k \downarrow}^\dag c_{\vec k' \uparrow} - {\rm H. c.} \bigr) 
\label{eq:eom}
\end{eqnarray}
Thus, computing the spin current injected at an interface between a PMI and a NM is reduced to computing 
the following correlation function \cite{Takahashi2010}: 
$j_{int}^s = \langle \hat j_{int}^s \rangle= - J_{eff}^2 {\rm Im} [U_{ret} (\delta\mu)]$.  
The correlation function $U_{ret} (\delta\mu)$ is conveniently computed by the analytic continuation 
of the imaginary-time correlation function with the bosonic Matsubara frequency $i \Omega_l \rightarrow \delta\mu + i\delta$ as 
$U(i \Omega_l) =  - \int_0^\beta d \tau e^{i \Omega_l \tau} \langle T_\tau A (\tau) A^\dag(0) \rangle$ 
with
$A(\tau) = \sum_{\vec q, \vec q', \vec k, \vec k'} a_{\vec q \downarrow}^\dag (\tau)  a_{\vec q' \uparrow} (\tau) 
c_{\vec k \uparrow}^\dag (\tau)  c_{\vec k' \downarrow} (\tau) 
$. 
Using the Wick theorem, this is reduced to 
\begin{equation}
U(i \Omega_l) = \!  \int_0^\beta \!\!\! d \tau e^{i \Omega_l \tau} \!\!\!\! \sum_{\vec q, \vec q', \vec k, \vec k'} \!\!\!\!
D_{\vec q \downarrow}(-\tau) D_{\vec q' \uparrow}(\tau) G_{\vec k \uparrow}(-\tau) G_{\vec k' \downarrow}(\tau), 
\label{eq:UiOmega}
\end{equation}
with its diagrammatic representation shown in Fig.~\ref{fig:cartoon} (c). 
Here, $G$ and $D$ are the normal Green's functions for conduction electrons and APs, respectively. 
Note that the anomalous Green's function $F$ does not appear even for the AFM SBs and this expression holds 
for all the cases considered here. 
This is because $F_{\vec q}$ involves the additional momentum dependence proportional to $\sum_\rho \sin q_\rho$, 
and its contributions disappear by the $\vec q$ integral. 
When the spin splitting is for APs in the PM region, the expression for the interface spin current, $j_{int}^s$, is not changed, 
but the sign is reversed as seen from the equation of motion for the spin, Eq.~(\ref{eq:eom}). 

Using the Matsubara Green's functions for conduction electrons $G_{\vec k \sigma} (i \omega_n) = (i \omega_n - \xi_{\vec k})^{-1}$ with $\omega_n = (2n+1) \pi T$
and APs $D_{\vec q \sigma} (i \nu_n)$ in Eq.~(\ref{eq:D1}),  
it is straightforward to obtain 
\begin{widetext}
\begin{equation}
j_{int}^s = - \pi J_{eff}^2 N_F^2(0) \!\!\!
\sum_{\xi_{\vec k}, \vec q, \vec q'} \!\! \!
[n_F(\xi_{\vec k}) - n_F(\xi_{\vec k} - \omega_{\vec q'} + \omega_{\vec q} + \delta\mu)]
[ n_B(\omega_{\vec q'} - \omega_{\vec q} - \delta\mu) \mp n_X(\omega_{\vec q'})]
[n_X(\omega_{\vec q}) - n_X(\omega_{\vec q} +  \delta\mu)]  ,
\label{eq:jint1}
\end{equation}
for the FM SBs and the AFM SFs. 
Here, $\xi_{\vec k}$ is the energy of a conduction electron with momentum $\vec k$. 
$X=B$ for FM SBs and $F$ for AFM SFs, and $n_{B/F}(x)=[\exp(x/T) \mp 1]^{-1}$ is the Bose-Einstein/Fermi-Dirac distribution function. 
As usual, we approximated the momentum integral for electrons in the NM region by the energy $\xi_{\vec k}$ integral near the Fermi level 
assuming the constant density of states $N_F(0)$.     
%
Typical  energy scales for $\omega_{\vec q}$ and $\delta\mu$ are much smaller than the electron Fermi energy, 
allowing the replacement of $n_F(\xi_{\vec k} + \alpha)$ by $n_F(\xi_{\vec k}) - \delta(\xi_{\vec k}) \alpha$. 
By further taking the limit of  $\delta\mu \to 0$, we arrive at 
\begin{eqnarray}
j_{int}^s =\pi J_{eff}^2 N_F^2(0) \delta\mu 
\sum_{\vec q, \vec q'} (\omega_{\vec q} - \omega_{\vec q'})  n_{X}' (\omega_{\vec q}) 
[n_B (\omega_{\vec q'}-\omega_{\vec q}) \mp n_{X}(\omega_{\vec q'})], 
\label{eq:jint3}
\end{eqnarray}
with $n'_X(\omega) = \partial n_X(\omega)/\partial \omega$. 
For the AFM SB case, a similar calculation leads to 
\begin{eqnarray}
j_{int}^s \!\! &= & \!\! - \pi J_{eff}^2 N_F^2(0) \!\!
\sum_{\xi_{\vec k}, \vec q, \vec q'} \! \! \sum_{{\scriptstyle s,s'}\atop{\scriptstyle = \pm 1}}\!\! 
\frac{\omega_{\vec q}+s \lambda}{2\omega_{\vec q}} \frac{\omega_{\vec q'}+s' \lambda}{2\omega_{\vec q}} \nonumber \\
&\times&\!\! 
[n_F(\xi_{\vec k}) - n_F(\xi_{\vec k} - s' \omega_{\vec q'} +  s \omega_{\vec q} + \delta\mu)]
[ n_B(s' \omega_{\vec q'} - s \omega_{\vec q} - \delta\mu) - n_B(s' \omega_{\vec q'})]
[n_B(s \omega_{\vec q}) - n_B(s \omega_{\vec q} + \delta\mu)]  .
\label{eq:jint4}
\end{eqnarray}
Similarly, by expanding $n_F(\xi_{\vec k}+\alpha)$ at $\xi_{\vec k}=0$ and taking the limit of $\delta\mu \to 0$, we obtain 
\begin{eqnarray}
j_{int}^s = \pi J_{eff}^2 N_F^2(0) \delta\mu 
\sum_{\vec q, \vec q'} \sum_{s,s'}  
\frac{\omega_{\vec q}+s \lambda}{2\omega_{\vec q}}
\frac{\omega_{\vec q'}+s' \lambda}{2\omega_{\vec q'}}
(s \omega_{\vec q} - s' \omega_{\vec q'}) n_{B}' (s \omega_{\vec q})
[n_B (s' \omega_{\vec q'}-s \omega_{\vec q}) - n_{B}(s' \omega_{\vec q'})] .  
\label{eq:jint5}
\end{eqnarray}

\end{widetext}

\begin{figure}[htbp]
\includegraphics[width=0.7\columnwidth, clip]{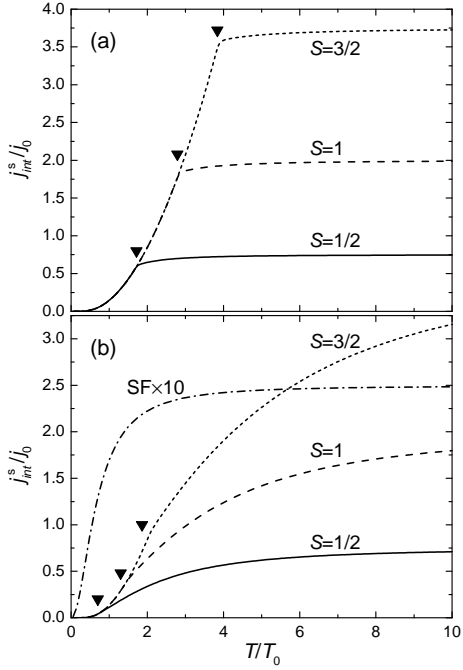}
\caption{Spin current injected at a MI/NM interface for FM cases (a) and AFM cases (b). 
$j_0 = \pi J_{eff}^2 N_F^2(0) \delta\mu/T_0$. 
$T_0 = J\chi$ for FM SBs and AFM SFs, while $T_0=J\Delta$ for AFM SBs. 
For SBs, solid lines are for $S=1/2$, broken lines $S=1$, dotted lines $S=3/2$. 
The SF result is shown by a dash-dotted line. 
Bose condensation occurs at a temperature $T_c$ indicated by a triangle, 
below which all curves overlap because the temperature is scaled by $\chi$ or $\Delta$. 
}
\label{fig:injection}
\end{figure}

In contrast to the magnon contribution discussed in Ref. \cite{Takahashi2010}, 
the above formulas involve additional Bose or Fermi factors as well as energy differences 
representing the exchange of spin between APs inside the PM region or electrons inside the NM region.  
This is because the time-reversal symmetry is maintained even below the Bose condensation temperature. 
As a consequence, for the FM case, the spin injection is proportional to $T^3$ at low temperatures. 
This can be easily seen as follows: 
First, momentum integrals $\sum_{\vec q}$ are replaced by energy integrals $\int_0 d \varepsilon D(\varepsilon)$ 
with some upper cutoff and the SB density of states $D(\varepsilon) \propto \sqrt{\varepsilon}$. 
Then, the energy is rescaled as $\varepsilon/T = \tilde \varepsilon$. 
At low temperatures, $\tilde \varepsilon$ integrals can be taken from 0 to infinity because of the Bose-Einstein distribution function, 
arriving at the leading term proportional to $T^3$ in Eq.~(\ref{eq:jint3}).  
By contrast, the $T^{3/2}$ behavior is predicted for FM insulator/NM interfaces by magnon contributions \cite{Takahashi2010} 
and similar behavior is experimentally confirmed for FM metal/NM interfaces \cite{Garzon2005}. 
Since the $T^{3/2}$ behavior is for the configuration where injected spin is parallel or antiparallel to the ordered moment, 
the current temperature dependence might be realized when the injected spin is perpendicular to the ordered moment.   

The numerical results for the spin current injection $j_{int}^s$ are presented in Fig.~\ref{fig:injection} (a) 
for the FM model given by Eq.~(\ref{eq:jint3}) and Fig.~\ref{fig:injection} (b) for the AFM model given by Eq.~(\ref{eq:jint5}). 
For the AFM model, we plot both SB results with different $S$ and a SF result with $S=1/2$. 
For the FM case, $j_{int}^s$ rapidly decreases with decreasing temperature below the magnetic transition temperature indicated by a triangle. 
This behavior is consistent with the analysis using a spin wave theory except for its power law. 
Importantly, this does not seem to depend on the sign of interaction, FM or AFM, the methodology, SB or SF, or even 
the presence of magnetic ordering. 
At high  temperatures, the results depend on the detail of the model and method, but the qualitative behavior is again rather universal; 
$j_{int}^s$ depends on temperature very weakly. 
For the AFM SF, $j_{int}^s$ is flat above its characteristic temperature, $\sim J\chi$, below which the AFM correlation is increased. 
The weak temperature dependence of $j_{int}^s$ comes from the competition between the factor $1/T$ originating from $\partial n_X(\omega)/\partial \omega$
and the energy window available for exchanging spin, i.e., the factor $(\omega_{\vec q} - \omega_{\vec q'})$, whose  contributions also increase with temperature. 
The temperature dependence for AFM SB is somewhat stronger than the other cases. 
This is because the pairing of SBs suppresses the spin fluctuation more efficiently. 

{\em Spin conductivity}. 
The performance of devices consisting of a MI connected with NM electrodes 
does not only depend on the spin current injection discussed in the previous section but also on how injected spins transfer inside the MI or  
from one electrode to the other electrode through the MI. 
This problem might require the treatment of nonequilibrium conditions including both NM regions and a MI region. 
However, this treatment is so far only available at very low temperatures \cite{Chen2013}. 
Here, instead we discuss the spin conductivity of bulk MIs. 
Combined with the spin injection discussed in the previous section, this should provide important insights into the performance of MI/NM systems. 

We start from deriving the bulk spin current by the equation of motion for 
$S_{\vec r}^z = a_{\vec r \uparrow}^\dag a_{\vec r \uparrow} -a_{\vec r \downarrow}^\dag a_{\vec r \downarrow}$. 
Under the MF approximation, the spin current on the bond ($\vec{r}$, $\vec{r} + \hat{x}$) is expressed as 
$j_x^s(\vec r) = \frac{J\chi}{4} (i a_{\vec r \uparrow}^\dag a_{\vec r + \hat x \uparrow} - i a_{\vec r \downarrow}^\dag a_{\vec r + \hat x \downarrow} + {\rm H.c.})$ 
for the FM SB and the AFM SF cases and 
$j_x^s(\vec r) = \frac{J\Delta}{4} (i a_{\vec r \uparrow} a_{\vec r + \hat x \downarrow} + i a_{\vec r \downarrow} a_{\vec r + \hat x \uparrow} + {\rm H.c.})$ 
for the AF SB case. 
In a momentum space, these spin current operators in the small-$\vec q$ limit become 
\begin{equation}
j_x^s (\vec{q}) = \frac{J\chi}{2} 
\sum_{\vec k} \sin k_x \Bigl(a_{\vec k \uparrow}^\dag a_{\vec k + \vec q \uparrow}
-a_{\vec{k}\downarrow}^\dag   a_{\vec k + \vec q \downarrow}\Bigr) 
\label{eq:j1}
\end{equation}
for the FM SBs and the AFM SFs and 
\begin{equation}
j_x^s (\vec{q}) = \frac{J\Delta}{2} i
\sum_{\vec k} \cos k_x \Bigl(a_{-\vec k +\vec q \downarrow} a_{\vec k \uparrow}
-a_{\vec{k}\uparrow}^\dag   a_{-\vec k - \vec q \downarrow}^\dag \Bigr) 
\label{eq:j2}
\end{equation}
for the AFM SBs. 

\begin{figure}[tbp]
\includegraphics[width=0.7\columnwidth, clip]{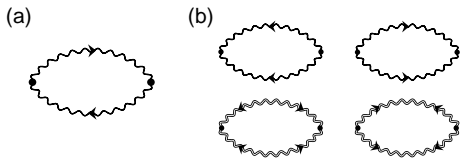}
\caption{Feynman diagram representing the spin conductance for  
(a) FM SBs and AFM SFs and (b) AFM SBs. 
Single wavy lines represent the normal propagators of APs, while double wavy lines are the anomalous propagators. 
Dots are the vertexes. }
\label{fig:bubble}
\end{figure}

The spin conductivity $\sigma (\omega)$ is obtained by using the standard Kubo formula \cite{Sentef2007}. 
For SB cases, care must be taken for the condensed components of SBs, which represent the ordered moments not superconductivity \cite{Schafroth1955}. 
While the condensed components could introduce novel spin superconductivity \cite{Takei2014a,Takei2014b}, 
this requires an additional theoretical or experimental setup, the planer spin anisotropy for FM or an external magnetic field for AFM, 
and the analyses on the dynamics of the condensed components, which is beyond the scope of our study. 
Thus, we focus on the contribution from the fluctuating components to the spin conductivity, 
which is given by the analytic continuation of 
$\sigma(i \Omega_l) = i \int_0^{1/T} d \tau e^{i \Omega_l \tau} \langle T_\tau j^s_x (\vec q, \tau) j^s_x (-\vec q,0) \rangle |_{\vec q \to 0} /i \Omega_l$ 
with $i \Omega_l \to \omega + i \delta$. 
For FM SBs and AFM SFs, 
the current-current correlation function 
$\chi (\tau) = \langle T_\tau j^s_x (\vec q, \tau) j^s_x (-\vec q,0) \rangle$  
has the well known form 
$\pm (J\chi/2)^2 \sum_{\vec k, \sigma} \sin^2 k_x D_{\vec k \sigma} (-\tau) D_{\vec k+ \vec q \sigma}(\tau)$. 
For AFM SBs, 
this is given by  
$
\chi (\tau) = 
(J \Delta/2)^2  \sum_{\vec k} \cos^2 k_x 
\times [ D_{-\vec k+\vec q \downarrow} (\tau) D_{\vec k \uparrow} (\tau) 
+ D_{\vec k \uparrow} (-\tau) D_{-\vec k-\vec q \downarrow} (-\tau) 
- F_{-\vec k}^* (-\tau) F_{\vec k+\vec q}^* (\tau) - F_{\vec k-\vec q} (-\tau) F_{\vec k} (\tau)] $. 
%
These correlation functions are diagramatically shown in Fig.~\ref{fig:bubble}. 
Without disorders and interactions, the conductivity consists of a delta function located at $\omega=0$ with the amplitude $D$, i.e., Drude weight. 
For FM SBs and AFM SFs, this is given by 
\begin{eqnarray}
D= - 2 \pi \biggl( \frac{J\chi}{2} \biggr)^2 \sum_{\vec k} \sin^2 \! k_x  n_X' (\omega_{\vec k}), 
\label{eq:d1}
\end{eqnarray}
and for AFM SB we find 
\begin{equation}
D= - \pi \biggl( \frac{J\Delta}{2} \biggr)^2 \!\! \sum_{\vec k, s=\pm1} \!\! 
\cos^2 \! k_x \Biggl( \frac{J \Delta \gamma'_{\vec k}}{\omega_{\vec k}} \Biggr)^2 
n_B' (s \omega_{\vec k}) . 
\label{eq:d2}
\end{equation}

\begin{figure}[htbp]
\includegraphics[width=0.7\columnwidth, clip]{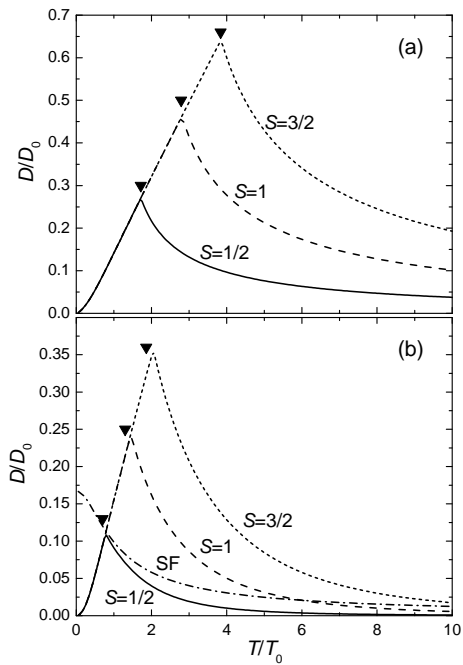}
\caption{Coefficient $D$ for the spin conductance $\sigma(\omega) = D \delta(\omega)$ as a function of temperature
for the FM cases (a) and the AFM cases (b). 
$D_0= \pi J\chi/2 $ for FM SBs and AFM SFs, while $D_0= \pi J \Delta/2$ for AFM SBs. 
Triangles indicate the magnetic transition temperatures, $T_C$ for FM and $T_N$ for AFM.}
\label{fig:conductance}
\end{figure}

Figure \ref{fig:conductance} shows the numerical results for $D$ for the FM model (a) and the AFM model (b). 
For the AFM model, both SB and SF results are shown. 
At high temperatures, $D$ increases with decreasing temperature for all the cases. 
But for SBs, $D$ is cutoff at the Bose condensation or magnetic ordering indicated by triangles because the fluctuating components decrease rapidly 
(the maximum of $D$ is located slightly above $T_c$). 
These $D$ become zero at zero temperature, which is consistent with the results using the spin wave theory \cite{Meier2003,Sentef2007}. 
Such a temperature behavior is analogous to that of the uniform magnetic susceptibility, maximizing or diverging at $T_c$, 
as magnetic systems become most susceptible against external (magnetic) perturbations. 
By contrast, the spin conductance does not diverge even for a FM. 
This is because the spin current operator and the magnetic moment have the different momentum dependence or the form factor, 
see Eqs. (\ref{eq:j1}) and (\ref{eq:j2}), and only thermally populated SBs contribute to the spin conductance.  
Within the MF approximation used here, AFM SFs behave as non-interacting fermions. 
Thus, the conductance shows the standard behavior due to the thermal broadening. 
At finite temperatures, $D$ decreases with increasing temperature roughly proportional to $(T/T_0)^{-2}$ in all the cases. 
In fact, $T_0$ is proportional to the MF order parameter, $\chi$ or $\Delta$, which should also have the $T$ dependence in principle. 
Since $\chi^2$ or $\Delta^2$ is related to $|\langle \vec S_{\vec r} \cdot \vec S_{\vec r'}\rangle|$, roughly proportional to $T^{-1}$, 
 $D$ is expected to be scaled as $T^{-3}$ at much higher temperatures than the ordering temperature. 

\begin{figure}[htbp]
\includegraphics[width=0.7\columnwidth, clip]{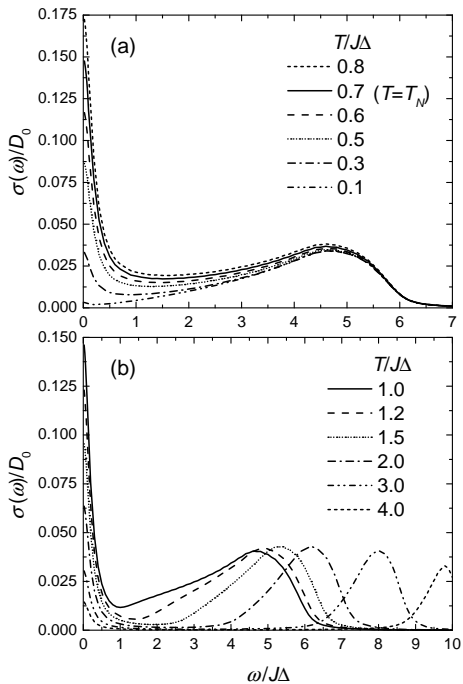}
\caption{The spin conductance $\sigma(\omega)$ for the AFM SB with $S=1/2$ as a function of $\omega$ at various temperatures indicated. 
AFM ordering occurs below $T_N =0.7 J\Delta$.
Here, the finite imaginary part of the SB self-energy is introduced as $\delta=0.1 J\Delta$.}
\label{fig:conductance_w}
\end{figure}

When the effect of disorder or the Gilbert dumping \cite{Gilbert2004} is to introduce a finite lifetime for APs, $\tau_{qp}$ ($qp$ stands for quasiparticle), 
$\sigma (\omega)$ becomes a Lorentzian as $\sigma(\omega) = D \tau_{qp}/\pi [ (\omega \tau_{qp})^2+1]$, and the dc conductivity becomes finite. 
For the AFM SB case, there appears additional continuum at finite frequencies extended up to $\omega \sim 2 \times \max (\omega_{\vec q})$ 
because of the two particle creation/annihilation processes, which is analogous to the result of the spin wave theory \cite{Sentef2007}. 
We computed $\sigma(\omega)$ for AFM SB with $S=1/2$ using the Green's functions Eq.~(\ref{eq:D2}) analytically continued to the real frequency as $i \nu_n \rightarrow \omega + i \delta$ 
with $\delta =1/2\tau_{qp}$ representing the finite self-energy. 
The results are shown in Fig.~\ref{fig:conductance_w}. 
The low-frequency Drude feature follows the temperature dependence of $D$ in Fig.~\ref{fig:conductance} (b). 
At temperatures above $T_N=0.7J\Delta$, the continuum moves to high frequencies because the gap amplitude in the SB excitation spectra increases. 

Within the theoretical model, the AP lifetime would be induced by the particle-particle interactions and beyond MF effects. 
The former might give a temperature dependence similar to that of the magnon lifetime due to the magnon-magnon interactions 
at low temperatures: 
inversely proportional to $\omega_{\vec q}^2 [\ln (\omega_{\vec q}/T)]^2$ for FM, 
and inversely proportional to $\omega_{\vec q}^2 T^3 (|\ln T|+a)$ with $a$ a numerical constant for AFM \cite{Harris1971}. 
The lifetime due to the beyond MF effects and that due to the particle-particle interactions at elevated temperatures have not been evaluated. 
But in analogy to the argument for doped Mott insulators \cite{Lee2006}, 
these are expected to be a decreasing function of $T$.  
Thus, the maximum near $T_c$ would remain in the spin conductance, but the sharp peak could broaden. 

Now we discuss the implication of our results to related experiments. 
Shiomi and Saitoh measured the inverse spin Hall voltage $V_{ISHE}$ of La$_2$NiMnO$_6$/Pt devices and 
found that $V_{ISHE}$  is peaked around $T_C$ for FM La$_2$NiMnO$_6$ which generates spin current by spin pumping  \cite{Shiomi14}. 
Meanwhile, Qiu {\it et al.} performed a similar experiment on Y$_3$Fe$_5$O$_{12}$/CoO/Pt devices, 
where AFM CoO is inserted as a spacer, and found that $V_{ISHE}$ is peaked around $T_N$ for CoO \cite{Qiu15}. 
In the latter experiment, it was argued that the enhancement of $V_{ISHE}$ comes from the temperature dependence of the spin conductance in CoO. 
Our results on the spin conductance strongly support this conjecture. 
The additional strong suppression of $V_{ISHE}$ at low temperatures might come from the temperature dependence of the spin injection at interfaces with AFM CoO. 
On the other hand, it was argued that a similar enhancement is due to the broadening of the microwave resonance shape as there is no spacer layer between 
the spin current generator  La$_2$NiMnO$_6$ and the detector Pt in Ref. \cite{Shiomi14}. 
Our results suggest that both the spin conductivity in La$_2$NiMnO$_6$ and the spin injection at the interfaces contribute to $V_{ISHE}$ 
especially at high temperatures as the generated spin current has to propagate inside La$_2$NiMnO$_6$ and from La$_2$NiMnO$_6$ to Pt.  
Qiu {\it et al.} also measured the dependence of $V_{ISHE}$ on the the microwave frequency $f$ for the spin pumping. 
At $T_N$, $V_{ISHE}$ is reduced by a factor of 4 from $f=4$ GHz to $8$ GHz. 
Assuming $8$ GHz corresponds to the inverse of the SB life time $\tau_{qp}$, one estimates $\delta = 1/2 \tau_{qp} \approx 0.01$ meV. 
Since the typical magnetic coupling is 10--100 meV, it is reasonable that only the Drude feature is detectable in the GHz range. 
In Ref. \cite{Wu2015}, Wu {\it et al.} report that the voltage response due to the PM SSE is proportional to $T^{-\alpha}$ with $\alpha \sim 3$--4. 
Our results indicate that a part of this temperature dependence is from the intrinsic temperature dependence of the spin conductivity in a PMI and the spin injection at the interface. 
However, detailed analyses for this experiment require even more information, such as the temperature dependence of order parameters, 
the lifetime of APs due to disorder, particle-particle interactions, and the interactions with phonons. 
%


As related subjects, there have appeared some theoretical proposals to detect novel electronic states of spin liquid systems by interfacing with conventional systems. 
References \cite{Chen2013,Chatterjee2015} proposed a similar experimental setup, the injection of spin current into a MI, 
to detect spinon or magnon excitation spectra inside the insulator. 
Their focus is the voltage dependence of the spin injection through a clean interface at low temperatures. 
On the other hand, our focus is rather the temperature dependence of the spin injection through a rough interface where momentum transfer is not conserved. 
Thus, it should be experimentally more easily accessible. 
Reference \cite{Norman2009} considered a junction consisting of a spin liquid insulator sandwiched by two FMs. 
They proposed that a spinon Fermi surface could induce a characteristic oscillatory behavior in the magnetic coupling between the two FMs 
as a function of the thickness of the spin liquid spacer. 
In our case, a magnetic coupling between NM and PMI is rather dynamical, the injection of spin current in or out of PMI. 
Further, we focused on the universal behavior of PMI described by both bosonic and fermionic APs, which do not necessarily have a Fermi surface. 

In contrast to metal- or semiconductor-based magnetic interfaces as widely discussed in the literature \cite{Zutic2002}, 
MI/NM interfaces do not necessarily accompany the charge accumulation inside a magnetic region when injecting or extracting spins. 
Instead, when the interface spin accumulation exists, 
the Lagrange multiplier or the local chemical potential, enforcing the local constraint $\sum_\sigma a_{\vec r \sigma}^\dag a_{\vec r \sigma}=2S$, 
may have to be fixed accordingly using techniques developed for inhomogeneous magnetic systems \cite{Okamoto2009,Ghosh2015}. 
One of the interesting directions is combining techniques developed in Refs. \cite{Zutic2002,Okamoto2009,Ghosh2015} and applying them to 
nanostructures with the spin injector, detector, and a magnetic/PM insulator; similar structures are originally proposed in Ref. \cite{Johnson1985}. 
Such nanostructures might be useful to measure the spin diffusion length in magnetic/PM insulators.

\section{Conclusion}
To summarize, we have presented rather simple descriptions to understand the spin current injection at a paramagnetic insulator/normal metal interface 
and the spin current transport inside a paramagnetic or magnetic insulator based on the Schwinger boson and Schwinger fermion mean-field theories. 
Both the spin current injection $j_{int}^s$ and spin current transport $D$ show rather universal behaviors irrespective of the sign of the magnetic coupling, 
FM or AFM, and methodology: 
$j_{int}^s$ depends on temperature rather weakly above magnetic transition temperature, 
while it decreases rapidly with decreasing temperature below the transition temperature. 
Similarly, $D$ decreases with increasing temperature above the transition temperature, 
and rapidly decreases with decreasing temperature below the transition temperature. 
These findings suggest that there is an optimal temperature window for the performance of a magnetic insulator/normal metal junction 
around the magnetic transition temperature of the constituent magnetic insulator. 
Such a device would potentially function even above the magnetic transition temperature. 
The current results support the conjecture of Ref. \cite{Qiu15}---%
the enhancement of $V_{ISHE}$ around $T_N$ is mainly due to the spin-current transport by thermal magnons inside an AFM spacer---%
and suggest that a similar mechanism contributes to the enhancement of $V_{ISHE}$ around $T_C$ of a FM insulator used for the spin current generator without a spacer \cite{Shiomi14}. 
It would be also interesting to analyze nonequilibrium situations in which a magnetic insulator is sandwiched by two normal metals \cite{Zutic2002} 
with different spin accumulations and temperatures. 
Our approach would provide some useful insights into the phenomena reported in Ref. \cite{Wu2015}. 
Finally, further experimental tests for the frequency-dependent spin conductance, as also predicted in Ref. \cite{Sentef2007}, would be desirable, 
but experimental techniques remain to be developed. 


\acknowledgements
S.O. thanks  A. Bhattacharya, S. Takahashi, N. Nagaosa, S. Maekawa, D. Xiao, D. Hou, Z. Qiu and E. Saitoh for fruitful discussions. 
The research by S.O. is supported by 
the U.S. Department of Energy,  Office of Science, Basic Energy Sciences, Materials Sciences and Engineering Division. 




\begin{thebibliography}{*}

\bibitem{Maekawa2006}{\it Concepts in Spin Electronics}, edited  by S. Maekawa (Oxford University Press, 2006). 

\bibitem{Dyakonov1971}M. I. Dyakonov and V. I. Perel, Phys. Lett. {\bf A35}, 459 (1971).
\bibitem{Hirsch1999}J. E. Hirsch, Phys. Rev. Lett. {\bf 83}, 1834 (1999).
\bibitem{Zhang2000}S. Zhang, Phys. Rev. Lett. {\bf 85}, 393 (2000).
\bibitem{Murakami2003}S. Murakami, N. Nagaosa, and S. C. Zhang, Science {\bf 301}, 1348 (2003).
\bibitem{Sinova2004}J. Sinova, D. Culcer, Q. Niu, N. A. Sinitsyn, T. Jungwirth, and A. H. MacDonald, Phys. Rev. Lett. {\bf 92}, 126603 (2004).

\bibitem{Saitoh2006}E. Saitoh, M. Ueda, H. Miyajima, G. Tatara, App. Phys. Lett. {\bf 88}, 182509 (2006). 

\bibitem{Bauer2012}G. E. W. Bauer, E. Saitoh, and B. J. van Wees, Nature Matter. {\bf 11}, 391 (2012). 

\bibitem{Uchida2008}K. Uchida, S. Takahashi, K. Harii, J. Ieda, W. Koshibae, K. Ando, S. Maekawa, and E. Saitoh, Nature (London) {\bf 455} 778 (2008). 

\bibitem{Uchida2010}K. Uchida, J. Xiao, H. Adachi, J. Ohe, S. Takahashi, J. Ieda, T. Ota, Y. Kajiwara, H. Umezawa, H. Kawai, G. E.W. Bauer, S. Maekawa, and E. Saitoh,
Nature Mater.  {\bf 9}, 894 (2010). 


\bibitem{Wang2014}H. Wang, C. Du, P. C. Hammel, and F. Yang, Phys. Rev. Lett. {\bf 113}, 097202 (2014). 

\bibitem{Kajiwara2010}Y. Kajiwara, K. Harii, S. Takahashi, J. Ohe, K. Uchida, M. Mizuguchi, H. Umezawa, H. Kawai, K. Ando,
K. Takanashi, S. Maekawa, and E. Saitoh, Nature (London) {\bf 464}, 262 (2010). 

\bibitem{Maekawa13}S. Maekawa, H. Adachi, K. Uchida, J. Ieda, and E. Saitoh, J. Phys. Soc. Jpn. {\bf 82}, 102002 (2013). 

\bibitem{Shiomi14}Y. Shiomi and E. Saitoh, Phys. Rev. Lett. {\bf 113}, 266602 (2014). 

\bibitem{Qiu15}Z. Qiu, J. Li, D. Hou, E. Arenholz, A. T. N'Diaye, A. Tan, K. Uchida, K. Sato, Y. Tserkovnyak, Z. Q. Qiu, and E. Saitoh, arXiv:1505.03926. 

\bibitem{Wu2015}S. M. Wu, J. E. Pearson, and A. Bhattacharya, Phys. Rev. Lett. {\bf 114}, 186602 (2015). 

\bibitem{Arovas1988}D. P. Arovas and A. Auerbach, Phys. Rev. B {\bf 38}, 316 (1988). 

\bibitem{Sinova2015}J. Sinova, S. O. Valenzuela, J. Wunderlich, C. H. Back, and T. Jungwirth, Rev. Mod. Phys. {\bf 87}, 1213 (2015). 

\bibitem{Zutic2002}I. {\v Z}uti{\'c}, J. Fabian, and S. Das Sarma, Phys. Rev. Lett. {\bf 88}, 066603 (2002). 

\bibitem{Takahashi2010}S. Takahashi, E. Saitoh, and S. Maekawa, J. Phys.: Conf. Ser. {\bf 200}, 062030 (2010). 

\bibitem{Garzon2005}S. Garzon, I. {\v Z}uti{\'c}, and R. A. Webb, Phys. Rev. Lett. {\bf 94}, 176601 (2005). 



\bibitem{Chen2013}C.-Z. Chen, Q.-f. Sun, F. Wang, and X. C. Xie, Phys. Rev. B {\bf 88}, 041405(R) (2013). 

\bibitem{Sentef2007}M. Sentef, M. Kollar, and A. P. Kampf, Phys. Rev. B {\bf 75}, 214403 (2007). 

\bibitem{Schafroth1955}M. R. Schafroth, Phys. Rev. {\bf 100}, 463 (1955). 

\bibitem{Takei2014a}S. Takei and Y. Tserkovnyak, Phys. Rev. Lett. {\bf112}, 227201 (2014). 
\bibitem{Takei2014b}S. Takei, B. I. Halperin, A. Yacoby, and Y. Tserkovnyak, Phys. Rev. B {\bf 90}, 094408 (2014). 

\bibitem{Meier2003}F. Meier and D. Loss, Phys. Rev. Lett. {\bf 90}, 167204 (2003). 

\bibitem{Gilbert2004}T. L. Gilbert, IEEE Trans. Mag. {\bf 40}, 3443 (2004). 

\bibitem{Harris1971}A. B. Harris, D. Kumar, B. I. Halperin, and P. C. Hohenberg, Phys. Rev. B {\bf 3}, 961 (1971). 

\bibitem{Lee2006}P. A. Lee, N. Nagaosa, and X.-G. Wen, Rev. Mod. Phys. {\bf 78}, 17(2006). 


\bibitem{Chatterjee2015}S. Chatterjee and S. Sachdev, Phys. Rev. B {\bf 92}, 165113 (2015).  

\bibitem{Norman2009}M. R. Norman and T. Micklitz, Phys. Rev. Lett. {\bf 102}, 067204 (2009). 



\bibitem{Okamoto2009}S. Okamoto, J. Phys.: Condens. Matter {\bf 21}, 355601 (2009). 
\bibitem{Ghosh2015}S. Ghosh, H. J. Changlani, and C. L. Henley, Phys. Rev. B {\bf 92}, 064401 (2015). 

\bibitem{Johnson1985}M. Johnson and R. H. Silsbee, Phys. Rev. Lett. {\bf 55}, 1790 (1985). 

\end{thebibliography}
\end{document}